\documentclass[sigconf,authorversion]{acmart}
\AtBeginDocument{%
  \providecommand\BibTeX{{%
    \normalfont B\kern-0.5em{\scshape i\kern-0.25em b}\kern-0.8em\TeX}}}

\acmConference[ICBBE 2022]{2022 9th International Conference on Biomedical and Bioinformatics Engineering}{November 10--13, 2022}{Kyoto, Japan}
\acmBooktitle{2022 9th International Conference on Biomedical and Bioinformatics Engineering (ICBBE 2022), November 10--13, 2022, Kyoto, Japan}
\acmPrice{15.00}
\acmDOI{10.1145/3574198.3574205}
\acmISBN{978-1-4503-9722-3/22/11}

\usepackage{amsmath} 
\usepackage{graphicx}
\usepackage{hyperref}
\usepackage{subcaption}
\usepackage{multirow}
\usepackage{multicol}
\usepackage{booktabs}
\usepackage{graphicx}
\usepackage{color}
\begin{document}

\title{Slim U-Net: Efficient Anatomical Feature Preserving U-net Architecture for Ultrasound Image Segmentation}

\author{Deepak Raina}
\authornote{Equal contribution.}
\authornote{Deepak Raina (Corresponding author) is a doctoral student at the Indian Institute of Technology, Delhi, India, and currently a SERB-Overseas Visiting Doctoral Fellow at Purdue University, USA. Email: deepak.raina@mech.iitd.ac.in, draina@purdue.edu.}
\email{deepak.raina@mech.iitd.ac.in}
\orcid{0000-0002-5863-6111}
\affiliation{%
  \institution{Indian Institute of Technology}
  \city{Hauz Khas}
  \state{Delhi}
  \country{India}
  }

\author{Kashish Verma}
\authornotemark[1]
\email{vermakashish888@gmail.com}
\affiliation{%
  \institution{Indian Institute of Technology}
  \city{Hauz Khas}
  \state{Delhi}
  \country{India}
}

\author{Sheragaru Hanumanthappa Chandrashekhara}
\email{drchandruradioaiims@gmail.com}
\affiliation{%
  \institution{All India Institute of Medical Sciences}
\city{Ansari Nagar}
  \state{Delhi}
  \country{India}
}


\author{Subir Kumar Saha}
\email{saha@mech.iitd.ac.in}
\affiliation{%
  \institution{Indian Institute of Technology}
  \city{Hauz Khas}
  \state{Delhi}
  \country{India}
}

\renewcommand{\shortauthors}{D. Raina, K. Verma, S.H. Chandrashekhara, S.K. Saha}

\begin{abstract}
We investigate the applicability of U-Net based models for segmenting Urinary Bladder (UB) in male pelvic view UltraSound (US) images. The segmentation of UB in the US image aids radiologists in diagnosing the UB. However, UB in US images has arbitrary shapes, indistinct boundaries and considerably large inter- and intra-subject variability, making segmentation a quite challenging task. Our study of the state-of-the-art (SOTA) segmentation network, U-Net, for the problem reveals that it often fails to capture the salient characteristics of UB due to the varying shape and scales of anatomy in the noisy US image. Also, U-net has an excessive number of trainable parameters, reporting poor computational efficiency during training. We propose a Slim U-Net to address the challenges of UB segmentation. Slim U-Net proposes to efficiently preserve the salient features of UB by reshaping the structure of U-Net using a less number of 2D convolution layers in the contracting path, in order to preserve and impose them on expanding path. To effectively distinguish the blurred boundaries, we propose a novel annotation methodology, which includes the background area of the image at the boundary of a marked region of interest (RoI), thereby steering the model's attention towards boundaries. In addition, we suggested a combination of loss functions for network training in the complex segmentation of UB. The experimental results demonstrate that Slim U-net is statistically superior to U-net for UB segmentation. The Slim U-net further decreases the number of trainable parameters and training time by $54\%$ and $57.7\%$, respectively, compared to the standard U-Net, without compromising the segmentation accuracy. The project page with source code is available at \href{https://sites.google.com/view/slimunet}{\color{blue}\textit{{https://sites.google.com/view/slimunet}}}.

\end{abstract}
\maketitle
\section{Introduction} \label{sec:intro}
Ultrasound (US) imaging is one of the most commonly used diagnostic modalities for a variety of clinical examinations, including bedside and point-of-care diagnosis, pre- and post-operative evaluation, and facilitating surgical interventions. Compared to other imaging modalities, such as Computed Tomography (CT) and Medical Resonance Imaging (MRI), the US is the least ionizing, inexpensive, portable and provides real-time feedback. In all of these examinations, accurate segmentation of anatomical structures is desirable for appropriate diagnosis \cite{noble2006ultrasound}. In particular, segmentation of the Urinary Bladder (UB) in pelvic view US images is used for discriminating the bladder shape, identifying diverticula, stones, malignant tumours and free fluid (blood) during trauma. In general, the segmentation is manually done by the sonographer while manipulating the US probe over the patient body.  However, the increasing demand for these examinations places a significant burden on the medical community \cite{rohaya2011medical}, as the diagnosis by the US requires a skilled sonographer to obtain the appropriate images. In densely populated nations such as India or during pandemics such as COVID-19, it is exceedingly challenging to meet the enormous demand due to the shortage of experts \cite{gates2020responding}. Therefore, it is desirable to automate the UB segmentation in US images.
\begin{figure}[ht]
    \centering
	\includegraphics[trim=0cm 0cm 0cm 0cm,clip,width=\linewidth]{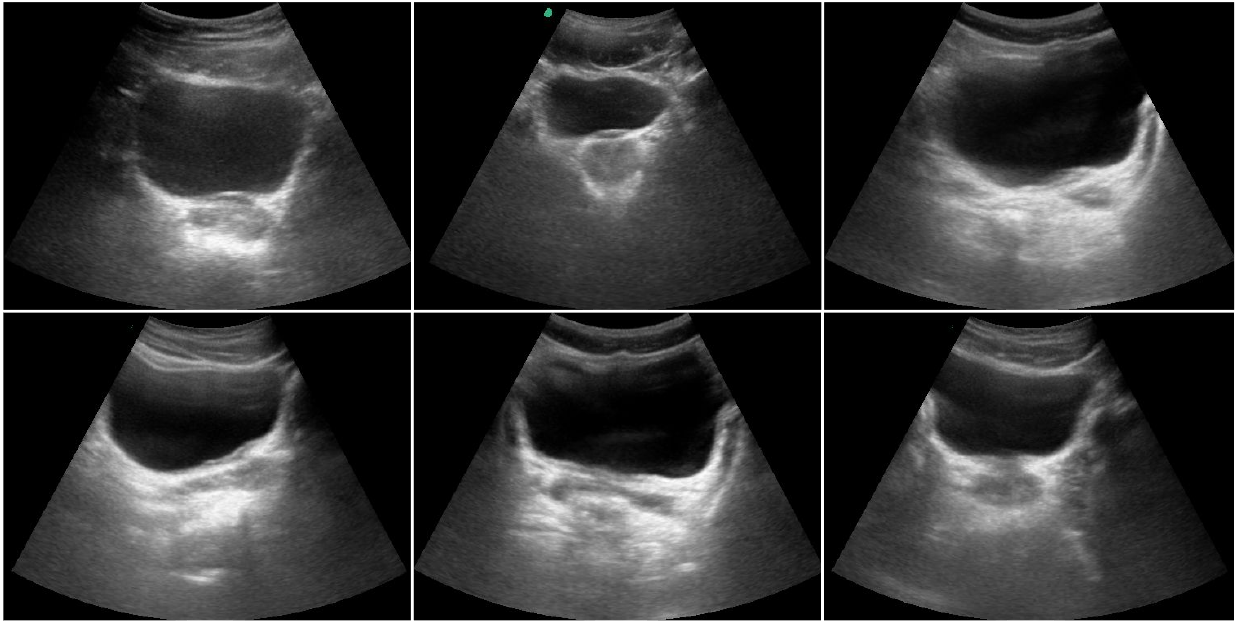}
	\caption{Ultrasound images dataset of male pelvic view. The urinary bladder in pelvic view shows large variability in shape and size along with blurred boundaries, thereby making the task of segmenting it quite challenging.}
    \label{fig:dataset}
\end{figure}

There exist several challenges to automatic US image segmentation. Unlike MRI or CT, US images exhibit spurious effects due to variations in anatomical structures, the existence of blurred image boundaries, shadowing artifacts, and artifacts caused by patient movement or handheld movement of the probe \cite{noble2016reflections}. Some of these effects can be seen in Fig. \ref{fig:dataset}, where UB is appearing in varying shapes and scales with indistinct boundaries, making the segmentation problem quite complex. In addition, manual segmentation by the sonographers causes inter- and intra-observer variability, thus restricting the reliability of UB diagnostics.

Numerous published articles have attempted to address the issue of US image segmentation for various procedures. In past, classical algorithms for ultrasound image segmentation methods have been presented using the clustering method \cite{chang2008segmentation}, watershed transform methods \cite{deka2006watershed, li2009medical}, Markov's random field method \cite{lihua2007segmentation}, active contours \cite{michailovich2007segmentation}, and statistical model-based methods \cite{sarti2005maximum}. In recent years, the huge success of Deep Convolutions Neural Networks (D-CNN) for natural image segmentation \cite{minaee2021image} has led to state-of-the-art results for medical image segmentation \cite{liu2021review}. Zhang et. al. \cite{zhang2016coarse} proposed a coarse-to-fine stacked fully convolution network (CFS-FCN) to incrementally segment the lymph nodes from ultrasound images. Their framework is comprised of two FCN-based modules, with the first FCN generating an intermediate segmentation map which is then utilized by the second FCN in conjunction with the raw image to produce the final lymph node segmentation. Wu et al. \cite{wu2017cascaded} introduced cascaded FCN (casFCN) for fetal ultrasound image segmentation to overcome the issues of boundary deficiency in images. In this work, the FCN model is integrated with the Auto-context scheme to stack a series of models that utilizes both the appearance features from images and the contextual features extracted by the preceding model. Mishra et al. \cite{mishra2018ultrasound} proposed a FCN model with attention to boundaries, where coarse resolution layers identify the object region from the background and fine resolution layers define the object's boundary. The framework is validated for blood region and lesion segmentation. However, the pooling operation of FCN, in general, has resulted in the loss of information for a pixel location in US images, degrading boundary details, and is therefore not conducive to medical image segmentation.

To overcome this issue, the encoder-decoder architecture has been developed to restore the spatial dimension and pixel location features by utilizing inverse operations, such as convolution-deconvolution and pooling-unpooling. Cunningham et al. \cite{cunningham2019ultrasound} used a encoder-decoder architecture, DeconvNet \cite{noh2015learning} for ultrasound segmentation of cervical muscle. Ronnerberger et al. \cite{ronneberger2015u} proposed a U-Net model consisting of an encoder-decoder with skip connections to incrementally adopt the long-range feature relationships. The encoder extracts high-level and low-level features, whereas the decoder reconstructs the image using skip connections. U-net performed exceptionally well on various biomedical image segmentation (BIS) tasks and won the ISBI 2015 competition. U-Net has numerous advantages, including (1) simultaneous use of global appearance features (handled by expanding path) and contextual features (handled by contracting path); (2) performs better for segmentation tasks even with a small training dataset; and (3) processes the entire image to create segmentation maps, which aids in preserving the comprehensive context of the input image, a significant advantage over patch-based segmentation methods \cite{milletari2016v, ronneberger2015u}. However, several limitations of the standard U-Net have been observed in segmenting medical imaging data \cite{kayalibay2017cnn, ibtehaz2020multiresunet}: (1) Difficulty in segmenting US images with anatomical structures of varying shapes and scales; (2) Majority of the fine-grained details are lost;  (3) Excessive number of trainable parameters that increases computation time during training and real-time testing. 

Several U-Net improvements were subsequently developed to solve its limitations and improve segmentation performance across various modalities. Lie et al. \cite{li2019automatic} modified the U-Net with deep connections (Dense U-Net) to segment the levator hiatus in ultrasound images of a female pelvic. The dense connections contributed to increasing the number of trainable parameters and feature reuse. Lin et al. \cite{lin2019semantic} proposed a semantic embedding and shape-aware U-Net model (SSU-Net) for eyeball segmentation, where they used a Signed Distance Field (SDF) instead of a binary mask as the label to learn the shape information and semantic embedding module to combine semantic features at coarser levels. Byra et al. \cite{byra2020breast} modified the U-Net utilizing selective kernels (SKU-Net) for breast mass segmentation in US images by replacing each convolution layer with an SK module with two branches, one of which generates feature maps using dilated convolutions and the other without dilation. The SKU-Net has proved to be effective at handling breast masses of varying sizes. Punn et al. \cite{punn2020inception} substituted the standard convolution layers with inception layers of Google-Net \cite{szegedy2015going} (Inception U-Net) to segment nuclei in microscopy cell images. Recently, they extended their model for breast cancer segmentation in ultrasound images using cross-spatial attention (CSA) block, which uses multi-scale information by concatenating multi-level encoder feature maps with corresponding decoder blocks to obtain better correlation and develop spatial attention feature maps \cite{punn2022rca}. Dunnhofer et al. \cite{dunnhofer2020siam} merged the U-net and Siamese framework (Siam-U-Net) for real-time tracking of femoral condyle cartilage in knee ultrasound images. Even though that numerous variations of U-Net have been proposed by adding convolutional layers or recurrent layers in skip connections, these modifications have either contributed more trainable parameters to the network or resulted in subpar performance for segmenting intricate anatomical features.
\\
\\
\textbf{Key contributions:} In order to address the complexity of UB for segmentation in US images, we propose a reshaped version of U-Net, termed Slim U-Net. The key contribution of this work can be summarized as follows:
\vspace{-0.1cm}
\begin{enumerate}
    \item We proposed a segmentation framework by reshaping the structure of standard U-Net \cite{ronneberger2015u} using lesser number of 2D convolution layers to preserve simple features extracted in the initial 2D convolution at each stage and imposing them on the expanding path. The reduction in convolution operations will help in avoiding the speckle noise-induced complexity of features, resulting in better segmentation.
    \item Considering the boundary of anatomical structure as an important clue for the segmentation model, we propose a new annotation methodology to consider the background area at the bounds of RoI to steer the model's attention towards the boundaries of UB. 
    \item We propose utilizing a combination of loss functions for training the network in order to extract fine-grained features of the bladder.
\end{enumerate}
\section{Dataset Collection and Curation} \label{sec:dataset}
\textbf{Dataset Collection}: The dataset consists of US images collected during the feasibility study of our in-house developed Telerobotic Ultrasound (TR-US) system \cite{tr-us, chandrashekhara2022robotic} with All India Institute of Medical Sciences (AIIMs), New Delhi (a public hospital and medical research university in India). The AIIMS Institute Ethics Committee approved the study (Ref. No. IEC-855/04.09.2020,RP-16/2020). The study was conducted from October 2021 to December 2021. Before the scanning operation, informed written consent was obtained from the volunteers, and their privacy was protected by anonymizing their identifiers in the image. The US dataset contains male pelvic view images, which consist of the urinary bladder, prostate and seminal vesicles. A sonographer records the images for each participant using a Sonosite M-TURBO ultrasound equipment with a C5-1 curved probe. The US machine's collected images were instantly transferred to the computer using Epiphan DVI2USB $3.0$ (Epiphan Video, Ontario, Canada) and saved in a .PNG format.
\\
\\
\textbf{Dataset Annotation:}
The deep-learning-based image segmentation algorithms rely heavily on expert-annotated images with marked regions of interest (RoI) using contours for optimal performance. There exist several annotation methods to effectively draw the contours in the images \cite{aljabri2022towards}. However, the manual annotation of US images is a time-consuming, labour-intensive, and expensive procedure. In our dataset, the boundary of the urinary bladder is required to be annotated in pelvic view US images. After careful evaluation of our US image dataset, it has been observed that the boundaries of the urinary bladder consist of a wide zone of pixels with gradually shifting grey levels (i.e dispersed) and open contours (i.e discontinued). Depending on the intricacy of the UB in US images, annotating a single image could take anywhere from minutes to hours. This would still be susceptible to annotation noise, which would affect the performance of the segmentation model. \cite{karimi2020deep}. 
\begin{figure}[ht]
    \centering
	\includegraphics[trim=0cm 5.5cm 0cm 0cm,clip,width=\linewidth]{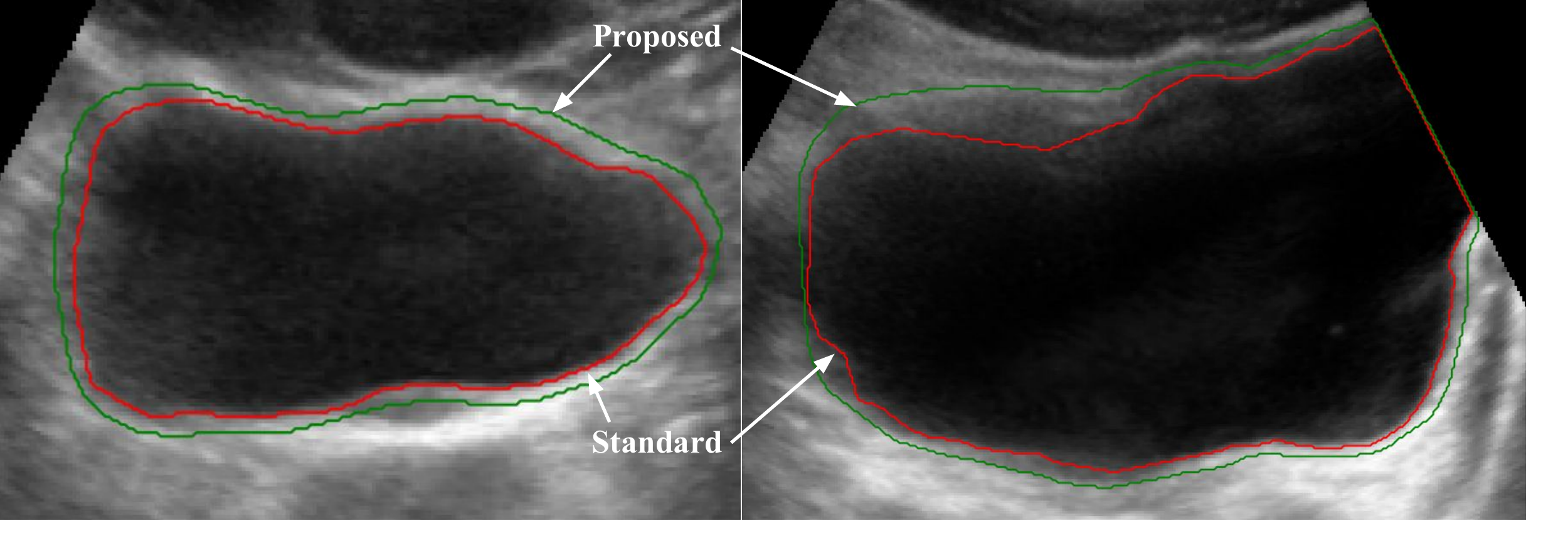}
    \caption{Visual comparison of annotation methodologies. Standard annotation methodology is represented by a red line, while the proposed annotation is represented by a green color line}
    \label{fig:annotation}
\end{figure}
\begin{figure*}[ht]
    \centering
	\includegraphics[trim=0cm 5.5cm 0cm 0cm,clip,width=\linewidth]{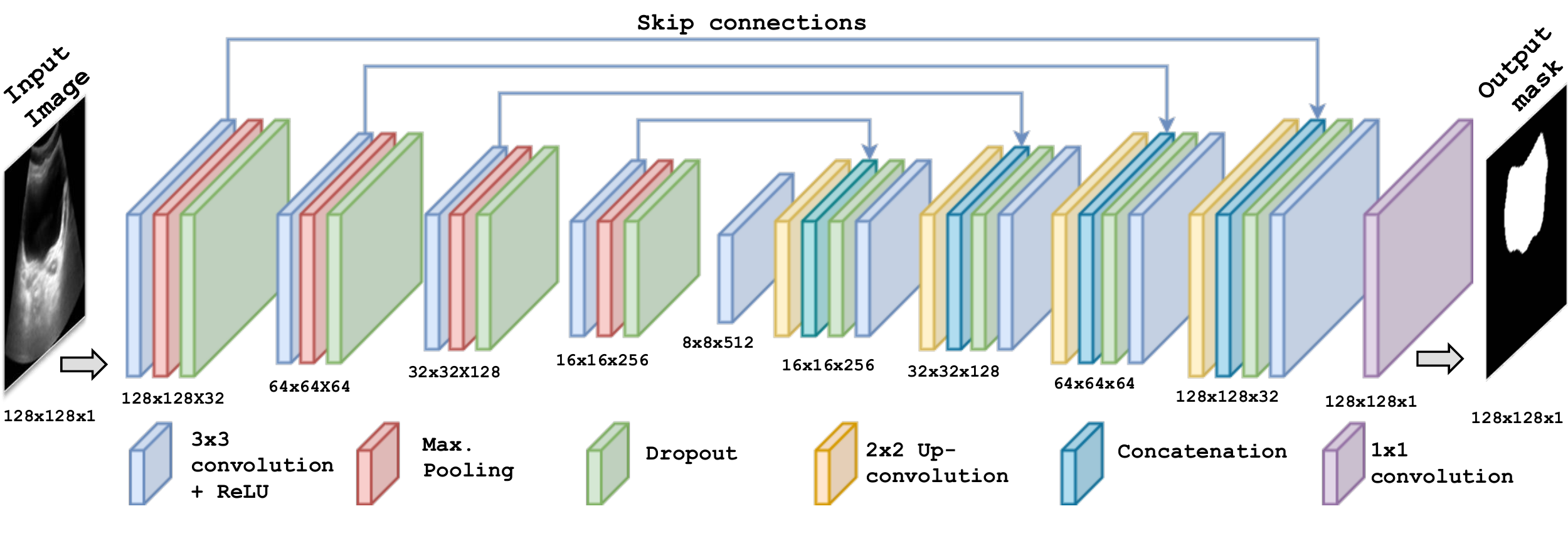}
	\caption{Architecture of Slim U-Net for urinary bladder (UB) segmentation in pelvic view US images. The left side represents the reshaped contracting path (encoder). At the bottom of each block is written $W \times H \times N$, where $H$ is the height, $W$ is the width, and $N$ is the number of channels.}
    \label{fig:arch}
\end{figure*}
To overcome these concerns, a novel method for annotating the UB boundary is provided. We have included the background area at the margins of marked RoI to steer the model's attention towards urinary bladder boundaries. The proposed annotation methodology is demonstrated in Fig. \ref{fig:annotation}. Our proposed annotation methodology will have the following advantages: (1) accurately captures the dispersed and discontinued boundaries definitions and makes the model robust to discriminating the UB from the background; (2) minimizes the labour, time and concentration required for annotating the images as the radiologist is only required to roughly mark the UB boundary. 

A radiologist with more than $15$ years of experience in abdomen radiology did the annotation of US dataset using SuperAnnotate (see \url{https://superannotate.com/}). A single polygon with multiple points is drawn in every image, spanning the irregular shape of UB using the proposed annotation technique. The polygons representing the ROI were then exported to a JSON file for all images in the dataset. Later, a python script has been used to generate corresponding ground truth masks with pixels labelled as $1$ representing the ROI and as $0$, representing the background. 
\\
\\
\textbf{Dataset Augmentation}
There exist several traditional and advanced methods for data augmentation to generate a diversity of images for the small dataset \cite{hussain2017differential}. The most common traditional methods used for images are scaling, cropping, flipping, rotation, adding noise and translation. These methods help to enhance the performance of CNNs and also addresses the issue of data scarcity, especially in medical imaging dataset. However, for medical images, it isn't necessary that the accuracy of CNNs will increase due to these data augmentation techniques \cite{hussain2017differential}. Especially for ultrasound images, it is recommended not to use rotation, scaling, cropping, adding noise and translation, as they will alter the characterization/physics of US images. The local features around the anatomical structures in ultrasound images appear according to the probe's standard field of view, position and orientation during scanning. As such, vertical flipping can not be applied due to the conical field of view of the US probe. However, horizontal flipping can be applied to mirror the anatomical structures, as they can appear on both sides of the cone during scanning. The horizontal flipping resulted in approximately doubling the dataset images. 
\\
\\
\textbf{Dataset Statistics}
After data collection, sorting and augmentation, we have $124$ images having a resolution of $504 \times 378$ pixels. All the ground truth masks were stored with the same file name of the corresponding raw image. Each image had three channels (Red, Green, and Blue), but we transformed them all to grayscale mode with a single channel. These images and ground truth masks were resized to $128 \times 128$ before training the model. 
\\
\\
\textbf{Dataset Splits}
Each dataset was split into training data and validation data with a $9:1$ ratio. Further, each model is validated using $10-$fold cross-validation metrics across the full dataset to assert how effectively the model generalises to new data. The cross-validation splits also ensured that the images from different subjects were present either in the training or test set.        


\section{Ultrasound Image Segmentation Methodology}

The basic architecture of the proposed US image segmentation model is inspired by U-net, which is a state-of-the-art segmentation network for BIS. We propose using a smaller number of 2D convolution layers in the standard U-Net's contracting path (encoder) and naming it as \textit{Slim U-Net}. Our hypothesis is to preserve simple features extracted in the initial 2D convolution in each stage and immediately impose them on the expanding path. In the case of Standard U-Net, two convolution layers prior to every max pooling extract more complex features that are not required for segmenting bladder in ultrasound images. Our proposed architecture performs better because a single layer before every max pooling layer extracts simple features to avoid the complexity imposed by speckle noise. Further, in pursuit of extracting more complexity may lead to learning the speckle noise in raw images. Thus, reducing the redundant convolutional layers and fine-tuning the model is a preferable option. As such, the Slim U-net would be able to handle the variability in the appearance of UB. Further, the reduction in the number of trainable parameters will decrease the computation time of the model without compromising the performance of the segmentation efficacy. Since the UB shape is quite complicated to segment, we have used multiple loss functions to train the network, so that the model would be able to capture fine-grained and coarse-grained characteristics of the dataset images. Note that the choice of a loss function is modality dependent and any combination of loss functions may not improve the network performance. The model will finally output the predicted mask of the given input US image.
\subsection{Model Architecture} \label{sec:model}
The proposed model architecture, Slim U-Net, is shown in Fig. \ref{fig:arch}. The network consists of one 2D convolution layer (Conv2D), followed by a Batch Normalization layer, Rectified Linear Unit (ReLU) as activation function, dropout as regularization layer, a max pooling layer (i.e down sampling) of $2 \times 2$ in the contracting path. Similarly for the expansive path, an Up-sampling layer of $2 \times 2$ was added after each Conv2D layer. Each Conv2D layer is padded convolution with kernel size as $3 \times 3$. Along the contracting path, the number of filters was doubled subsequently with each convolutional layer. The vice versa of this pattern was replicated in the expanding path, ending up with the same number of filters in the last $3 \times 3$ convolutional layer of expanding path. 
Another noteworthy feature in Slim U-net is a dropout layer after max pooling in the contracting path and after the concatenation layer in the expanding path. Dropout is an important hyper-parameter of neural networks and helps in preventing over-fitting on a small dataset. Our hyper-parameter selection experiments revealed that segmentation efficiency declines if the dropout rate is increased to $0.25$ or above using an Adam optimizer's default learning rate of $0.001$.

\subsection{Loss function} \label{sec:loss}
A loss function is an essential component of a medical image segmentation model and is used for penalizing the network during training if there is a divergence between the predicted pixel label and the ground truth label for each pixel. In the literature \cite{ma2021loss}, loss functions are classified into four categories as Region-based Loss, Boundary-based loss, Compound Loss and Distribution-based loss. Each loss function has its benefits and drawbacks, and its use is specific to segmentation tasks in image datasets. However, the commonly used segmentation losses for medical imaging are the cross-entropy and the soft dice score. Cross-entropy measures logarithm values of predicted probabilities and corrected probabilities of each pixel in the image. For the problems based on binary image segmentation, binary cross-entropy works well on pixel-level classification  \cite{jadon2020survey}. The Binary Cross-Entropy (BCE) loss function is mathematically expressed as follows:


\begin{equation}
    \boldsymbol{L}_{BCE} = -\boldsymbol{y}_{t} log(\boldsymbol{y}_{p})-(1-\boldsymbol{y}_{t}) log(1 - \boldsymbol{y}_{p})
\end{equation}
where $\boldsymbol{y}_{t}$ and $\boldsymbol{y}_{p}$ represents ground truth mask and predicted mask of the given US image, respectively. Another widely used loss function in medical image segmentation is the Dice coefficient (DC), which measures the overlap between the predicted image and the ground truth image. Consequently, DC consider the information both locally and globally, which is quite useful for achieving higher accuracy. Originally, the DC was developed for binary data and can be expressed as follows:
\begin{equation}
   DC = \frac{2\left | \boldsymbol{A} \bigcap \boldsymbol{B} \right |}{\left | \boldsymbol{A} \right |+\left | \boldsymbol{B} \right |}
\end{equation}
where $\left | \boldsymbol{A} \bigcap \boldsymbol{B} \right |$ represents the common elements between sets $\boldsymbol{A}$ and $\boldsymbol{B}$, and $\left | \boldsymbol{A} \right |$ represents the number of elements in set $\boldsymbol{A}$ (and likewise for set  $\boldsymbol{B}$). The loss function from DC is formulated for image segmentation tasks as follows:
\begin{equation}
    \boldsymbol{L}_{DC} = 1-\left ( \frac{2 \boldsymbol{y}_{t} \boldsymbol{y}_{p} + \boldsymbol{s}} {\boldsymbol{y}_{t} + \boldsymbol{y}_{p} + \boldsymbol{s}} \right)
\end{equation}
where $\boldsymbol{s}$ is smoothing constant, which will speed up the optimization \cite{smoothConstant} and prevent the division by zero in fraction.

There is another metric similar to the Dice coefficient i.e. Jaccard Index (JI), widely known as Intersection over Union (IoU). Jaccard index gives the ratio of intersection between the predicted mask and ground truth mask over the union of these two. This form of the Jaccard index is not differentiable. To address this issue, an approximation of the Jaccard index is used as a loss function, as follows:
\begin{equation}
 \boldsymbol{L}_{JI} = \boldsymbol{1} -  \frac{\boldsymbol{y}_{t} \boldsymbol{y}_{p} + \boldsymbol{s}} {\boldsymbol{y}_{t} + \boldsymbol{y}_{p} - \boldsymbol{y}_{t} \boldsymbol{y}_{p} + \boldsymbol{s}}
\end{equation}
For segmentation of the urinary bladder in our dataset, we have compared three combinations of the above three loss functions, defined as $ \boldsymbol{L}_{D}$, $\boldsymbol{L}_{DJ}$ and $\boldsymbol{L}_{DJB}$. 
\begin{gather}
\boldsymbol{L}_{D} = \boldsymbol{L}_{DC} \\
\boldsymbol{L}_{DJ} = \boldsymbol{L}_{DC} + \boldsymbol{L}_{JI} \\
\boldsymbol{L}_{DJB} = \boldsymbol{L}_{DC} + \boldsymbol{L}_{JI} + \boldsymbol{L}_{BCE}
\end{gather}
Our exhaustive study on the loss functions revealed that using $\boldsymbol{L}_{DJB}$ i.e. a combination of BCE, DC and JI loss gives the best accuracy in comparison to the other two loss functions, denoted by $\boldsymbol{L}_{D}$ and $\boldsymbol{L}_{DJ}$.
\section{Results and Discussions}

\subsection{Implementation Details}
The model has been implemented using Python $3.8$ and Keras $2.7.0$. The training and testing of the model have been done on a Dell workstation with NVIDIA Quadro RTX $5000$ with $16$GB memory.  All the images in the dataset and their corresponding masks were resized to $128 \times 128$ each. The input image is also converted to grey-scale mode. The optimizer used is Adam with a learning rate of $0.001$. The batch size used for training is $4$. The dropout rate was set to $0.125$ for all layers in the model. The model was trained with $50$ epochs with three types of callbacks, including Model checkpoint, Early stopping and Reduce learning rate on plateau. After analyzing learning curves, we realized that $50$ epochs are enough to compare two architectures as the best fit condition never went beyond $45$ epochs. Early stopping criteria have the patience of $10$ epochs. The Reduce learning rate on plateau callback was used to reduce the learning rate by a multiplication factor of $0.1$ with the patience of $5$ epochs if there is stagnation in validation loss. The lower bound for reducing the learning rate was set to $0.00001$. To avoid over-fitting on a limited US dataset, we used horizontal flips to augment the images in our dataset, as explained in Section \ref{sec:dataset}.
In the model network, the first $3 \times 3$ convolutional layer in the contracting path has $32$ number of filters. For other layers, the number of filters was picked as mentioned in the model architecture in section \ref{sec:model}. To normalize the output of each convolutional layer, batch normalization was performed before the ReLU activation layer. Finally, the sigmoid activation function was calculated over the last layer for pixel-level labelling to generate the output segmentation mask.
\subsection{Evaluation Metrics} \label{sec:evaluation}
In order to evaluate the performance of the proposed model for urinary bladder segmentation in pelvic view US images, we employed five metrics, namely, Precision (P), Recall (R), F1 Score (F1), Dice Coefficient (DC) and Intersection over Union (IoU). F1 score is a commonly used metric for image segmentation and is given by the harmonic mean of pixel precision and pixel recall. If the value of these metrics is close to $100$ (i.e. $100\%$), it implies that the output mask accurately overlaps with the ground truth mask. We also noted the training time per epoch to validate the model's computational efficiency. To assert the generalizability of the model on the dataset, we reported our results on $10$-fold cross-validation test (Mean$\pm$Standard deviation) and the fold with best performance, termed as best fold in this paper.
   

\subsection{Segmentation results for our dataset}
\begin{figure}
		\centering
		\includegraphics[trim=0cm 0cm 0cm 0cm,clip,width=0.85\linewidth]{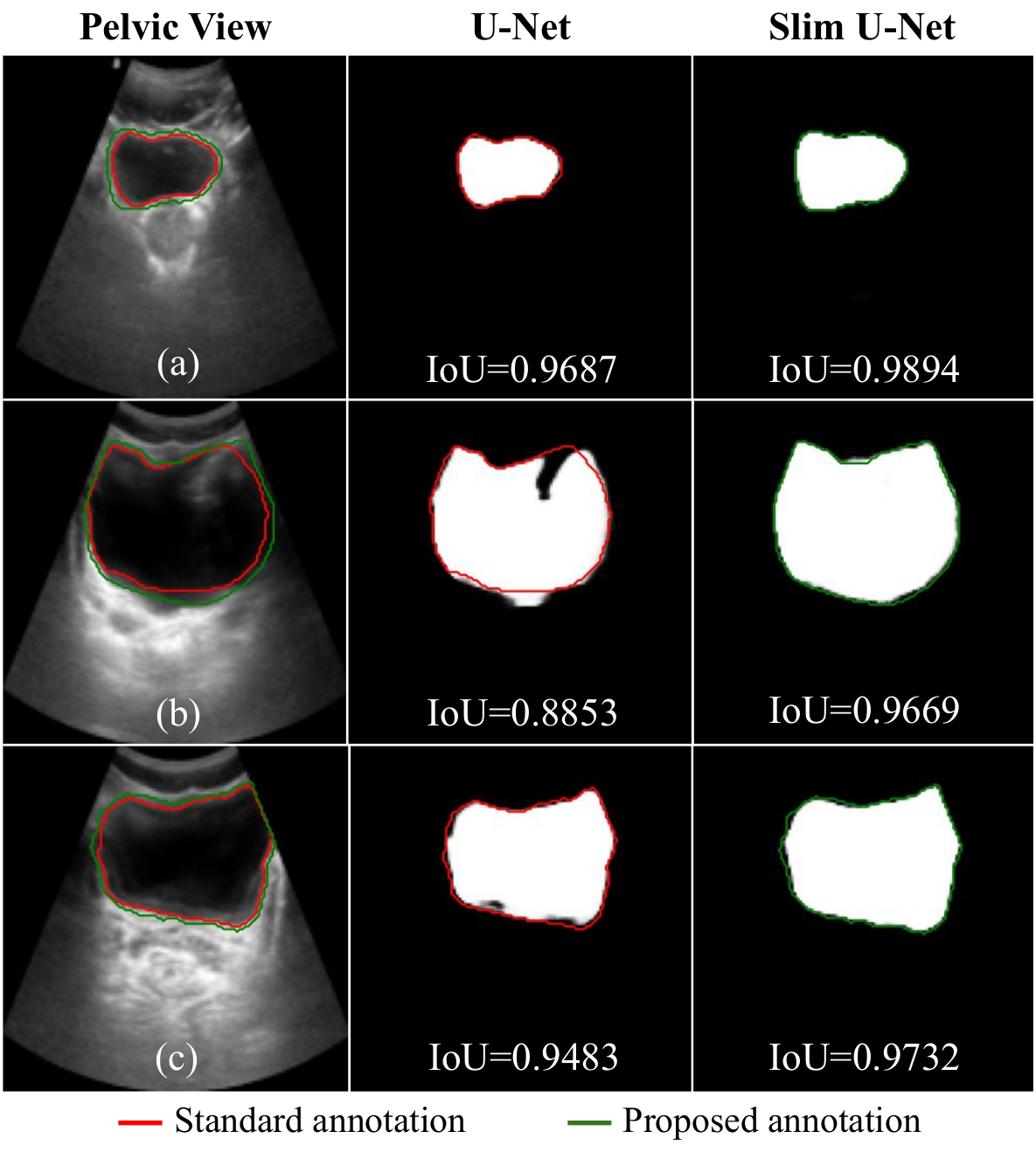}  
		\caption{The urinary bladder appears (a) at a smaller scale than regular anatomy, (b) with blurred boundaries, and (c) with regular anatomy but surrounded by prostate. Slim U-Net aids in finding salient regions by preserving global features and paying close attention to UB borders, an essential segmentation characteristic, and reports a higher IoU.} 
  \label{fig:teaser}
\end{figure}

\textbf{Qualitative evaluation:} We illustrate the urinary bladder segmentation results by comparing the results of Slim U-net with proposed annotation to those of standard U-Net with standard annotation, as shown in Fig. \ref{fig:teaser}. The proposed method successfully captures the discontinuous and deficient UB boundaries in the noisy regions of the image. The substantial variation in urinary bladder shape and scale is also inferred successfully. Hence, the proposed Slim U-net guided by our novel annotation methodology along with the multi-loss function is an effective method for ultrasound image segmentation.
\\
\\
\textbf{Quantitative evaluation:} We evaluate the proposed model using the evaluation criteria outlined in Section \ref{sec:evaluation}. The values of these evaluation metrics for the 10-fold cross-validation and the best fold are reported in Table \ref{tab:ablation123}. The best validation set scores on pelvic view US image dataset are: $98.9\%$ Precision, $98.75\%$ Recall, $98.82\%$ F1, $98.69\%$ DC and $97.42\%$ IoU. In $10$-fold cross-validation, the scores are: $98.60\pm0.36$ Precision, $97.98\pm0.74$ Recall, $98.29\pm0.34$ F1, $98.10\pm0.37$ DC and $97.29\pm0.71$ IoU, which shows the generalizability of the proposed Slim U-Net across the dataset.

\subsection{Ablation study}
\begin{figure}[ht]
		\centering
		\includegraphics[trim=0cm 0.5cm 8cm 0cm,clip,width=\linewidth]{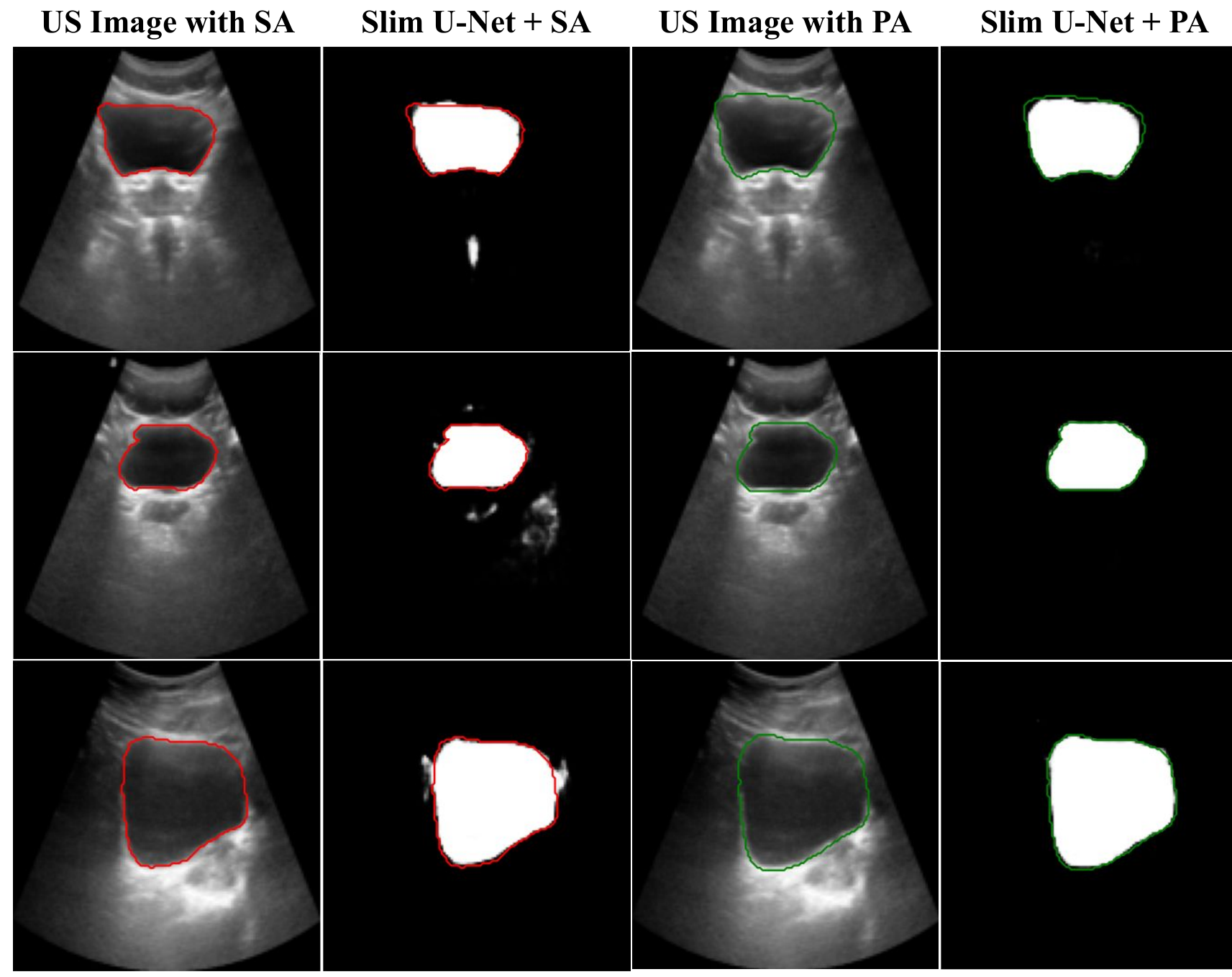}
		\caption{Qualitative evaluation of the ablated version of Slim U-net with Standard Annotation (SA) and Proposed Annotation (PA). The model with SA get influenced by the presence of prostate and noise in the image, while model with PA pinpoints the discriminating boundaries  and correctly predicts the mask.}
		\label{fig:ablation-annotation}
\end{figure}

\begin{figure}[ht]
		\centering
		\includegraphics[trim=0cm 1.5cm 9.5cm 0cm,clip,width=\linewidth]{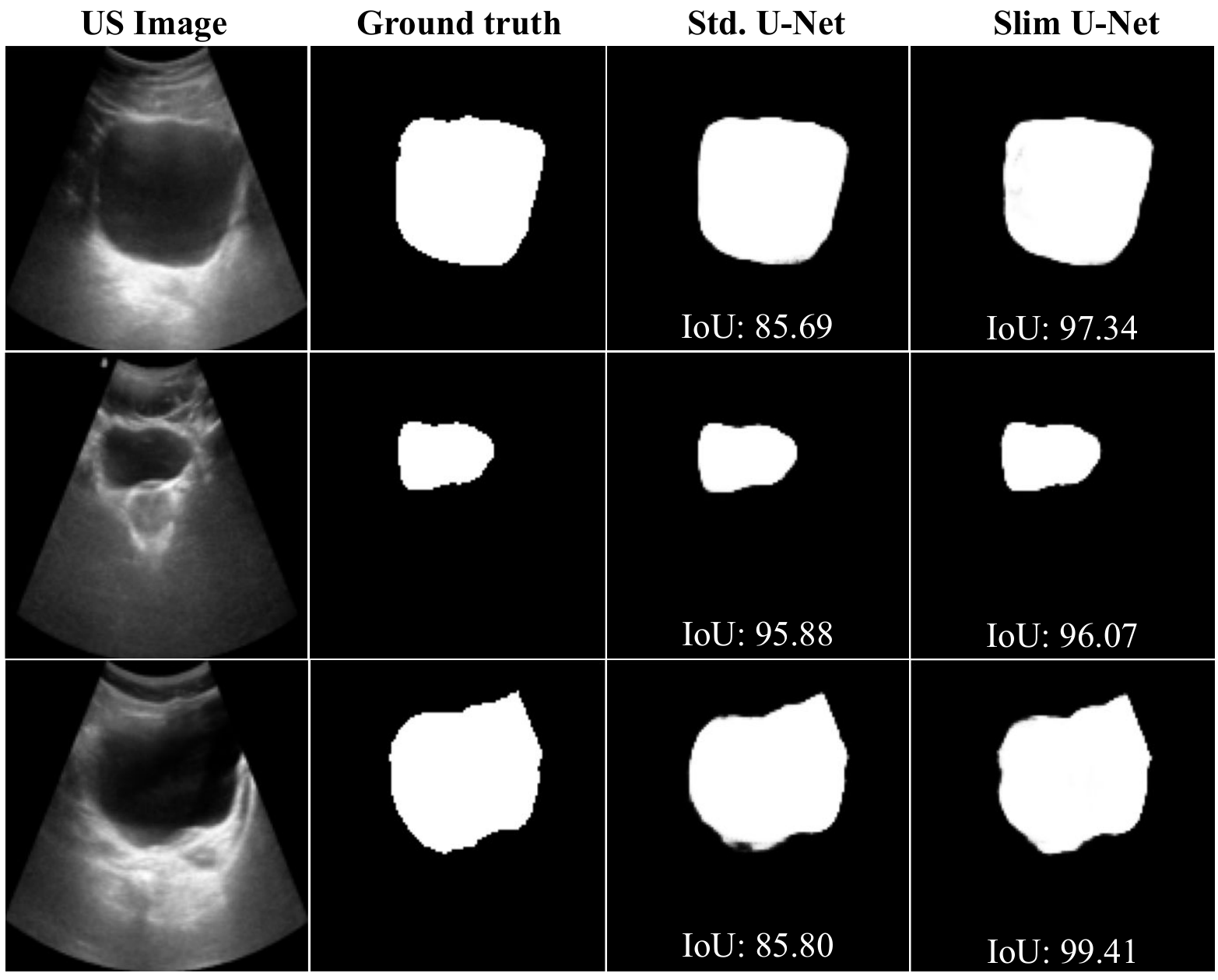}
		\caption{Comparison of predicted mask and IoU for ablated version of Slim U-net for different model architecture with best selected loss function ($\boldsymbol{L}_{DJB}$). Note that the Slim U-net outperforms the Standard (Std.) U-Net on IoU by $2-15\%$ for these three samples.}
		\label{fig:comparison}
\end{figure}

\begin{figure}[ht]
		\centering
		\includegraphics[trim=0cm 8.5cm 1cm 0cm,clip,width=\linewidth]{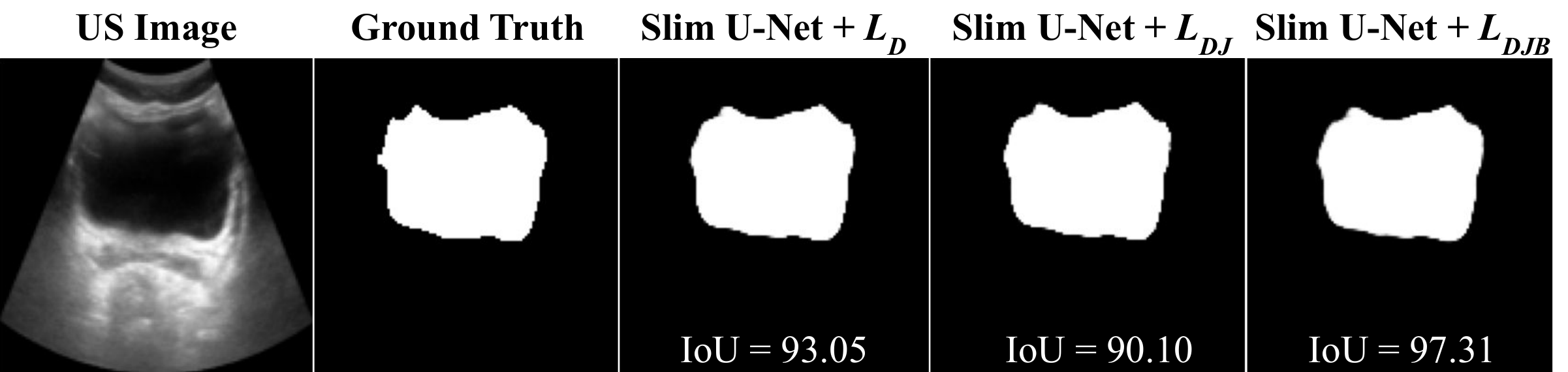}
		\caption{Comparison of predicted mask and IoU for ablated version of Slim U-net using different combination of loss function.}
		\label{fig:ablation-loss}
\end{figure}

We performed multiple ablation studies of the proposed Slim U-net to get better insight into the modifications proposed in the annotation methodology, network structure and loss function.
\\
\\
\textbf{Choice of Annotation:}
To validate the choice of proposed annotation, we compared the results of Slim U-net with the proposed annotation (Slim U-Net+PA) and with the standard annotation methodology (Slim U-Net+SA). In the case of standard annotation, the RoI boundary is marked to enclose the UB while excluding the background region as shown in Fig. \ref{fig:annotation}. The results of the ablation study are demonstrated in Table \ref{tab:annotate} and Fig. \ref{fig:ablation-annotation}. 
\begin{table}[H]
  \caption{Comparison of segmentation performance of the proposed Slim U-net on standard and proposed annotation.}
  \label{tab:annotate}
  \resizebox{\linewidth}{!}{\begin{tabular}{ccccccccccc}
    \toprule
    \multirow{2}{*}{\textbf{Annotation}} & \multicolumn{5}{c}{\textbf{Best Fold}} &  \multicolumn{5}{c}{\textbf{Cross Validation}} \\
    \cmidrule(lr){2-6}\cmidrule(lr){7-11}
    ~ & \textbf{P} & \textbf{R} & \textbf{F1} & \textbf{DC} & \textbf{IoU }& \textbf{P} & \textbf{R} & \textbf{F1} & \textbf{DC} & \textbf{IoU} \\
    \midrule
    Standard & $98.42$ & $98.83$ & $98.63$ & $98.50$ & $97.04$ & $98.79$ & $97.17$ & $97.96$ & $97.80$ & $95.72$ \\
    ~ & ~ & ~ & ~ & ~ & ~ & $\pm0.33$ & $\pm1.89$ & $\pm0.90$ & $\pm0.92$ & $\pm1.73$\\
    Proposed & $98.90$ & $98.75$ & $98.82$ & $98.69$ & $97.42$ & $98.60$ & $97.98$ & $98.29$ & $98.10$ & $97.29$\\   
    ~ & ~ & ~ & ~ & ~ & ~ & $\pm0.36$ & $\pm0.74$ & $\pm0.34$ & $\pm0.37$ & $\pm0.71$\\   
    \bottomrule
\end{tabular}}
\end{table}
\begin{table*}[ht]
    \centering
    \caption{Statistical analysis of ablated versions of the proposed Slim U-net for different model architecture and loss functions using the proposed annotation. The proposed model, Slim U-Net with $\boldsymbol{L}_{DJB}$ loss function significantly outperforms it's ablated versions on the test set and 10-fold cross validation test.}
    \resizebox{\linewidth}{!}{\begin{tabular}{cccccccccccc}
    \hline
        \multirow{2}{*}{\textbf{Model}} & \multirow{2}{*}{\textbf{Loss}} & \multicolumn{5}{c}{\textbf{Best Fold}} &  \multicolumn{5}{c}{\textbf{Cross Validation}} \\
        \cmidrule(lr){3-7}\cmidrule(lr){8-12}
        ~ & ~ & \textbf{P} & \textbf{R} & \textbf{F1} & \textbf{DC} & \textbf{IoU} &  \textbf{P} & \textbf{R} & \textbf{F1} & \textbf{DC} & \textbf{IoU}\\ 
        \hline
        Std. U-Net & \multirow{2}{*}{$\boldsymbol{L}_{D}$}  & $96.09$ & $98.13$ & $97.09$ & $97.1$ & $94.37$ & $42.35\pm45.5$ & $66.84\pm46.56$ & $45.08\pm42.82$ & $45.14\pm42.7$ & $39.56\pm41.88$ \\
        Slim U-Net & ~ & $98.51$ & $97.83$ & $98.16$ & $98.04$ & $96.17$ & $98.51\pm0.45$  & $97.83\pm0.86$ & $98.16\pm0.43$ & $98.04\pm0.54$ & $96.17\pm1.0$ \\
        \hline
        Std. U-Net & \multirow{2}{*}{$\boldsymbol{L}_{DJ}$} & $95.45$ & $96.78$ & $96.11$ & $96.11$ & $92.53$ & $10.94\pm30.02$ & $17.71\pm37.53$ & $12.0\pm30.49$ & $12.27\pm37.37$ & $10.88\pm28.99$\\ 
        Slim U-Net & ~ & $98.57$ & $97.74$ & $98.15$ & $98.04$ & $96.17$ & $98.57\pm0.43$ & $97.74\pm0.91$ & $98.15\pm0.44$ & $98.04\pm0.56$ & $96.17\pm1.07$\\ \hline
        Std U-Net & \multirow{2}{*}{$\boldsymbol{L}_{DJB}$} & $98.76$ & $98.56$ & $98.66$ & $98.54$ & $97.13$ & $98.04\pm0.66$ & $97.83\pm0.74$ & $97.93\pm0.41$ & $97.7\pm0.44$ & $95.52\pm0.84$\\ 
        \textbf{Slim U-Net} & ~ & $\boldsymbol{98.90}$ & $\boldsymbol{98.75}$ & $\boldsymbol{98.82}$ & $\boldsymbol{98.69}$ & $\boldsymbol{97.42}$ & $\boldsymbol{98.60}\pm\boldsymbol{0.36}$ & $\boldsymbol{97.98}\pm\boldsymbol{0.74}$ & $\boldsymbol{98.29}\pm\boldsymbol{0.34}$& $\boldsymbol{98.10}\pm\boldsymbol{0.37}$ & $\boldsymbol{97.29}\pm\boldsymbol{0.71}$\\ \hline

    \end{tabular}}
    \label{tab:ablation123}
\end{table*}
For the first US image (top row) in Fig. \ref{fig:ablation-annotation}, the model trained using standard annotation (SA) masks has segmented the prostate as UB. For the image in the second row, it predicts the UB pixels on the basis of low-level features. And the image in the third row fails to capture the clear boundaries of UB. However, the model trained using our proposed annotation masks shows promising results in all three samples. Thus, it is validated that the features learned on dispersed edges of the anatomical structures increase the segmentation accuracy in US images.
\\
\\
\textbf{Choice of model architecture:}
The proposed Slim U-net is compared with the standard U-Net to demonstrate the effectiveness of the proposed modification in network structure. The results are summarized in Table \ref{tab:ablation123} and illustrated using the US images in Fig \ref{fig:comparison}. Standard U-Net is the architecture proposed in \cite{ronneberger2015u}, which performs a block of two $3 \times 3$ convolutional layers before each down-sampling and up-sampling layer through contracting and expanding path. And the output of the convolutional block is also used in skip connections at different stages of the path. However, to standardize the ablation study, parameters such as input image size (i.e. $128 \times 128$), number of filters and batch size are kept similar for both the architectures. Note that we have also used the proposed annotation methodology in both standard (Std.) and proposed U-Net. From Table \ref{tab:ablation123}, it can be noted that Slim U-Net outperforms Std. U-Net on P, R, F1, DC and IoU by $0.14\%$, $0.19\%$, $0.16\%$, $0.15\%$ and $0.30\%$, respectively. Similar improvements have also been noticed for other loss functions and 10-cross validation tests. Also learning curves depicted that learning with Std. U-Net was overfitting for other loss functions. Thus, it is evident that our proposed model preserves the anatomical appearance features of UB with less number of contractions, reporting the highest IoU for segmenting it from male pelvic view US images.
\\
\\
\textbf{Choice of loss function:}
The proposed loss function is validated by comparing the performance of Slim U-Net by training it with various loss functions described in Section \ref{sec:loss}. The results are summarized in Table \ref{tab:ablation123} and illustrated using the US images in Fig \ref{fig:ablation-loss}. As reported in Table \ref{tab:ablation123}, the Slim U-net with proposed loss function, $\boldsymbol{L}_{DJB}$, reported the highest values for evaluation metrics and outperforms its ablated version with $\boldsymbol{L_{D}}$ on P, R, F1, DC and IoU by $0.39\%$, $0.93\%$, $0.67\%$, $0.66\%$ and $1.28\%$, respectively. Similar improvements have also been noticed over Slim U-net with $\boldsymbol{L}_{DJ}$. The results in Fig. \ref{fig:ablation-loss} shows that the Slim U-net with $\boldsymbol{L}_{DJB}$ achieves highest IoU of $97.31$, while its ablated version with loss function, $\boldsymbol{L}_{DJ}$ and $\boldsymbol{L}_{D}$,  achieves $90.10$ and $93.05$, respectively. We also noticed the improvement in the performance of standard U-net with the proposed loss function, which further validates our choice of loss function for the segmentation of urinary bladder in male pelvic view US images.
\subsection{Comparison of Computational efficiency}
We demonstrated the computational efficiency of the proposed Slim U-net by comparing the training time per epoch with the standard U-Net. The results are given in Table \ref{tab:freq}. The number of trainable parameters in the standard  U-net is $45.51\%$ more than the proposed Slim U-net. Due to the increase in the number of trainable parameters, standard U-Net reports low computational efficiency and takes  $42.30\%$ more training time per epoch than the proposed Slim U-Net. However, both architectures have approximately similar testing times. Thus, the proposed Slim U-net is computationally efficient without compromising the segmentation accuracy on a highly variable dataset of pelvic US images. 
\begin{table}[ht]
  \caption{Comparison of trainable parameters, training time per epoch (in sec) and testing time per image (in sec)}
  \label{tab:freq}
  \begin{tabular}{ccc}
    \toprule
    \textbf{Model} & \textbf{Standard U-Net} &  \textbf{Slim U-Net}\\
    \midrule
    {Trainable parameters} & $8,635,809$ & $4,705,377$\\
    {Training time per epoch} & $9.55$ & $5.51$ \\
    {Testing time} & $1.015$ & $1.015$ \\
    \bottomrule
\end{tabular}
\label{tab:compute}
\end{table}
\section{Conclusion}
We addressed the challenge of urinary bladder segmentation in male pelvic view ultrasound images. We presented a form of U-Net, Slim U-Net, with fewer 2D convolutions in the contracting path, which enables the network to retain the simple anatomical appearance features learnt in the contracting path and impose them on the expanding path, in order to deal with the enormous variety of UB in US images. We offer a unique annotation strategy and multiple loss functions to make our network resilient to indistinct UB boundaries and noise in the US image. The extensive experiments show that the Slim U-Net combined with proposed annotation and multiple loss functions improves the performance in comparison to standard U-Net. Moreover, the ablation analysis of the proposed framework validated the effect of model components on segmentation performance. Our future work will aim to study the effectiveness of the proposed algorithm for assisting the manual scanning by a sonographer as well as automatic scanning using robotic ultrasound proposed in \cite{tr-us}. We will also validate the generalizability of the model by applying Slim U-net for segmentation of other organs like kidney, gall bladder, liver, and prostate which pose challenges similar to UB in US images.
\begin{acks}
The first author gratefully acknowledges the Ministry of Education (MoE), Government of India, for providing the Prime Minster's Research Fellowship (Ref: F.No.35-5/2017-TS.I) to carry out his doctoral research program at Indian Institute of Technology, Delhi.
\end{acks}

\bibliographystyle{ACM-Reference-Format}
\bibliography{references}










\end{document}